\documentclass[%
 aps,prl,reprint,
 amsmath,amssymb,
floatfix
]{revtex4-2}

\usepackage{graphicx}
\usepackage{dcolumn}
\usepackage{bm}

\begin{document}

\preprint{APS/123-QED}

\title{Measurement of the $2$S$_{1/2}-8$D$_{5/2}$ transition in hydrogen}

\author{A. D. Brandt}

\author{S. F. Cooper}%
\author{C. Rasor}
\author{Z. Burkley}
 \altaffiliation[Now at ]{Dep. Physik, ETH Z\"{u}rich }
\author{D. C. Yost}
\email{Dylan.Yost@colostate.edu}

\affiliation{%
Department of Physics, Colorado State University, Fort Collins, CO 80523
}%

\author{A. Matveev}
 
\affiliation{Russian Quantum Center,  Skolkovo, Moscow 143025, Russia}%

\date{\today}

\begin{abstract}
We present a measurement of the hydrogen $2$S$_{1/2}-8$D$_{5/2}$ transition performed with a cryogenic atomic beam. The measured resonance frequency is $\nu=770649561570.9(2.0)$ kHz, which corresponds to a relative uncertainty of $2.6\times10^{-12}$. Combining our result with the most recent measurement of the $1$S$-2$S transition, we find a proton radius of $r_p=0.8584(51)$~fm and a Rydberg constant of $R_\infty=10973731.568332(52)$ m$^{-1}$. This result has a combined 3.1~$\sigma$ disagreement with the CODATA 2018 recommended value. Possible implications of the discrepancy are discussed.
\end{abstract}

\maketitle

Spectroscopy of hydrogen, the simplest element, was pivotal in the development of quantum theory and now plays a crucial role in the determination of fundamental constants and precision tests of fundamental theory \cite{Karshenboim2005, Karshenboim2007a, Mohr2016}. Due to its simplicity, theoretical calculations of the energy levels of hydrogen can be made with high accuracy, and deviations from theoretical predictions could suggest the presence of new physics \cite{Karshenboim2002, Safronova2018,  Tu2004, Karshenboim2010, Jones2019}.

The energy levels of hydrogen can be described by the following expression, given by \cite{Beyer2017, Grinin2020},
\begin{equation}
    E_{n,l,j}=hcR_\infty\left (-\frac{1}{n^2}+f_{n,l,j}(\alpha,\frac{m_e}{m_p},...)+\delta_{l,0}\frac{k_N}{n^3}r_p^2 \right).
   \label{hydrogen energy states}
\end{equation}
The first term, which depends only on the principle quantum number $n$, is the gross structure contribution from nonrelativistic quantum theory. The second term accounts for quantum electrodynamics (QED) and relativistic recoil corrections \cite{Mohr2016,Horbatsch2016}. The last term is the leading correction to the $S$ states arising from the finite size of the nucleus, where $r_p$ is the root-mean-square charge radius of the proton. By measuring two hydrogen transitions, the Rydberg constant, $R_\infty$, and $r_p$ in Eqn.~(\ref{hydrogen energy states}) can be determined. While $f_{n,l,j}$ is a function of other physical constants, such as the fine-structure constant and electron-to-proton mass ratio, each of these have been measured with sufficient precision in other experiments to not limit the determination of $R_\infty$ and $r_p$. Therefore, assuming that the QED corrections are accurately applied, a consistent extraction of $R_\infty$ and $r_p$ from spectroscopic measurements is expected.  

Tension arose when the value of $r_p$ obtained from muonic hydrogen was compared to the value determined from hydrogen spectroscopy and electron-proton elastic scattering data. This led to the ``proton-radius puzzle'' \cite{Pohl2010, Antognini2013a, Pohl2013, Carlson2015}. This discrepancy has spurred further interest in precision spectroscopy on hydrogen, and several new results have been recently published  \cite{Beyer2017,Fleurbaey2018,Bezginov2019,Grinin2020}. 

The CODATA 2014 recommended $r_p$ value, which is historically significant in the discussion of the proton-radius puzzle, is strongly influenced by previous measurements of the two-photon $2$S$-8$S/D transitions \cite{Biraben1986,DeBeauvoir2000a}. While remeasuring any of these transitions produces data relevant to the proton radius puzzle, the 2S$_{1/2}$-8D$_{5/2}$ transition is particularly attractive as it possesses the largest two-photon matrix element. Therefore, we have remeasured the $2$S$_{1/2}-8$D$_{5/2}$ transition and report a value that has a three-times smaller uncertainty as compared with the previous best measurement \cite{DeBeauvoir2000a}.

Our measurement has benefited from several technological advances. Improvements in laser technology afford us both the optical frequency comb to determine absolute optical frequencies, and the ability to optically excite a sufficient population of hydrogen atoms to the metastable 2S state, as in \cite{Beyer2017}. By generating metastable atoms optically via the two-photon 1S$_{1/2}$-2S$_{1/2}$ transition, instead of by electron bombardment, we are able to perform spectroscopy on a cryogenic beam of hydrogen atoms, reducing velocity-dependent systematics and allowing for the selective population of the $2$S$_{1/2}^{F=1}$ hyperfine manifold. Therefore, optical production of an appreciable flux of metastable hydrogen represents a substantial improvement over previous measurements of this transition. We have also directly characterized the velocity distribution, reducing uncertainty in velocity-related systematics \cite{Cooper2020}. Additionally, our apparatus provides a geometrically-constrained interaction between the hydrogen atoms and the spectroscopy light. 

\begin{figure*}[t]
\includegraphics[width=0.9\linewidth]{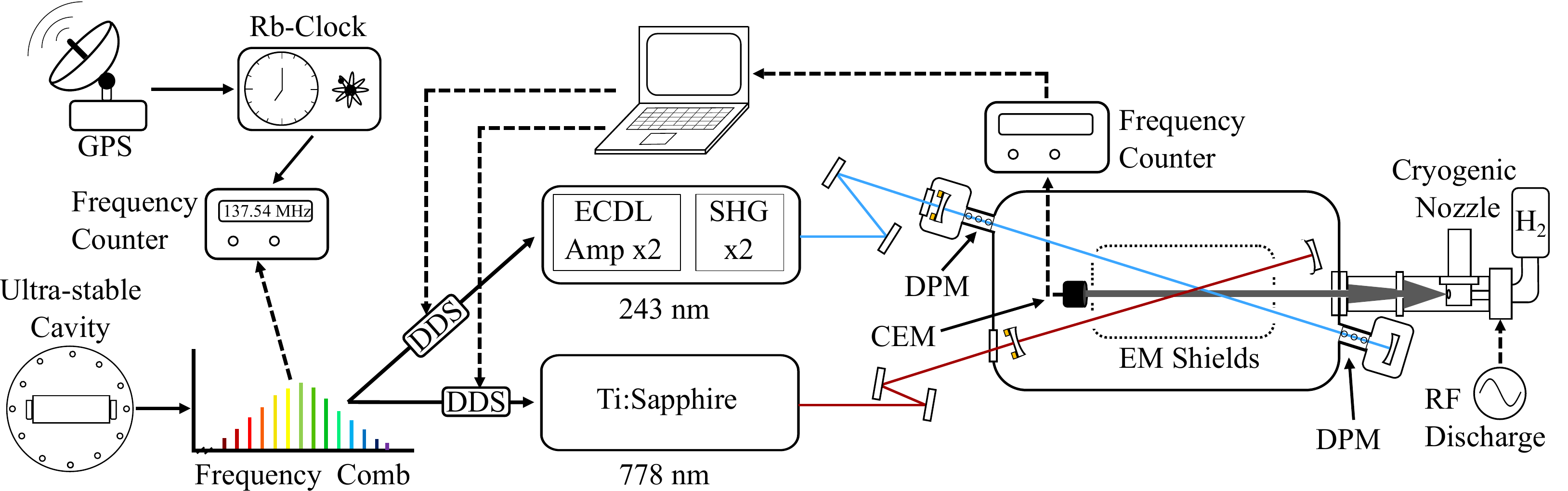}
\caption{Schematic of experiment. The repetition rate of the coherent frequency comb referenced to an ultra-stable optical cavity is counted by a GPS-disciplined Rubidium time standard. The 243 nm and 778 nm lasers are phase-locked to the comb, with the beat frequency set by a direct digital synthesizer (DDS). The Ti:Sapphire laser is a Coherent-899 vertically oriented ring laser with a modified piezoelectric controlled mirror to increase the frequency locking bandwidth. ECDL: Extended cavity diode laser. SHG: Second-harmonic generation.  DPM: Differential pumping manifolds. CEM: Channel electron multiplier. EM Shields: A pair of coaxial magnetic shields within a Faraday cage.}
\label{exp schematic}
\end{figure*}

Our experimental apparatus has not been fully described elsewhere. We generate a cryogenic beam of hydrogen by disassociating molecular hydrogen in a microwave discharge followed by a cryogenic nozzle \cite{Cooper2020}. At a distance of 1.5 m away from the nozzle, the atoms interact with 243 nm radiation to generate $2$S$_{1/2}^{F=1}$ population. The 243~nm radiation \cite{Burkley2019} is enhanced in an in-vacuum optical cavity \cite{Cooper2018} at a $6^{\circ}$ angle from the atomic beam. The UV cavity mirrors are kept in a $\sim200$ mTorr oxygen environment to prevent UV degradation, and a pair of differential pumping manifolds separates the mirror chambers from the spectroscopy region. The metastable 2S atoms then travel 15 cm before intersecting with 778 nm spectroscopy light from a Coherent-899 Ti:Sapphire ring laser, which is also enhanced in an in-vacuum optical cavity at a $6^{\circ}$ angle from the atomic beam.  The linearly polarized 778 nm radiation excites the 2S$_{1/2}-$8D$_{5/2}$ transition. Population in the 8D$_{5/2}$ state will rapidly decay -- predominantly to the 2P state which will then decay to the 1S ground state.  Therefore, by driving the 2S$_{1/2}-$8D$_{5/2}$ transition, we effectively quench the metastable population. The entire spectroscopic volume is within a pair of concentric magnetic shields, which is itself inside of a Faraday cage to mitigate Zeeman shifts and DC Stark shifts. The remaining metastable population not quenched by the 778 nm radiation is detected 15 cm past the $2$S$_{1/2}-8$D$_{5/2}$ interaction volume by a channel electron multiplier. The absolute frequencies of the 243 nm and 778 nm radiation are determined by phase-locking the lasers to a coherent Er-fiber optical frequency comb whose repetition rate is continuously counted by a GPS-disciplined, Rb time base and whose $f_0$ beat note is phase locked to an RF synthesizer. See Fig.~\ref{exp schematic} for a schematic of our experimental apparatus.

The UV-enhancement cavity is about 1.8 m long and is constructed with a pair of 1 m radius of curvature mirrors -- the effective $1/e^2$ beam radius is approximately $150$ $\mu$m at the intersection with the atomic beam. Between 1\% and 10\% of atoms in the excitation volume are driven to the $2$S$^{F=1}_{1/2}$ state. The spectroscopy light is also enhanced in an in-vacuum optical cavity to drive appreciable population to the 8D$_{5/2}$ state and the optical power is measured by continuously monitoring the cavity transmission with a photodiode. 
 
A direct digital synthesizer (DDS) sets the beat frequencies between the comb and the other laser systems -- the DDS itself is referenced to a 10 MHz signal from a GPS-disciplined, Rb time standard. To verify that our absolute frequency calibration is accurate, we locally counted a cesium-referenced 5 MHz signal at the NIST WWVB station in Fort Collins, Colorado with the Rb-time base and frequency counter, finding an accuracy within the expectations of a GPS-disciplined oscillator \cite{Lombardi2008} (a fractional accuracy of $\approx 5\times 10^{-13}$).  

To scan the $2$S$_{1/2}-8$D$_{5/2}$ transition, a set of 25 evenly spaced frequencies in a 3 MHz span around the resonance are chosen. This set of 25 frequency points is randomly sequenced. For each scan of the line, we fit the measured lineshape with
\begin{equation} 
    \mathcal{F}= A\hspace{1 pt}\text{exp}[-\alpha(\mathcal{L}_2(\nu_c,\gamma)+\mathcal{L}_3(\nu_c,\gamma))],
    \label{fit function}
\end{equation}
where $\mathcal{L}_i$'s correspond to  Lorentzian functions of appropriate relative magnitude centered on the $F=i$ hyperfine manifold, and $\{\alpha,\gamma,\nu_c\}$ are fit parameters for the amplitude of the metastable decrease, linewidth, and effective centroid frequency, respectively. Further details on the fitting function and other possible lineshape distortions (including quantum interference \cite{Udem2019}) can be found in the supplementary material (SM).

We have developed numeric lineshape models, which take into account the geometry and dynamics of the atom-light interaction, AC and DC Stark effects, velocity distribution, repopulation of the metastable state, and the second-order Doppler shift (see SM). The numeric models are primarily used to quantify the DC Stark effect, but are also used to verify the accuracy of fitting lineshapes with Eqn.~(\ref{fit function}). 

\begin{figure} [t]
    \centering
    \includegraphics[width=0.99\linewidth]{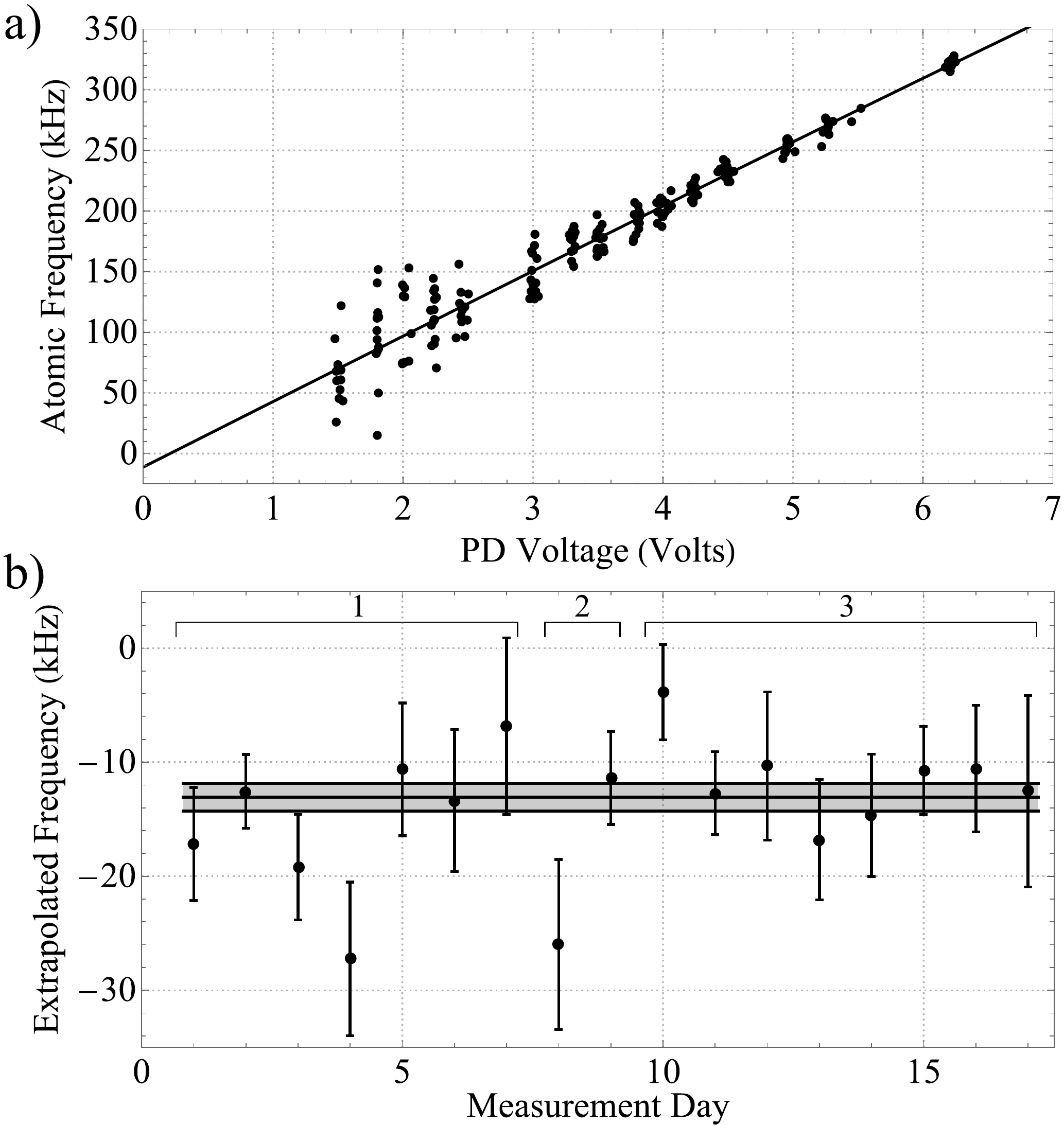}
    \caption{a) Example AC extrapolation data and cubic fit. b) Extrapolated zero-field frequencies for each measurement day with overall mean and statistical uncertainty overlaid. Data acquisition on the $2$S$_{1/2}-8$D$_{5/2}$ transition was collected in batches grouped by date, which are indicated with batch numbers 1, 2, and 3. Frequencies in both a) and b) are relative to the value listed in \cite{Kramida2010}.} 
    \label{extrapolation and scatter}
\end{figure}

The AC Stark shift is our leading systematic, with typical shifts in the range of 50-300 kHz. To properly account for this effect, we drive the resonance with a range of intensities, quenching between 15\% and 60\% of the metastable population, and extrapolate to zero laser power.  This extrapolation is similar to previous measurements of the 2S-$n$S/D transitions \cite{Schwob1999, DeBeauvoir1997, DeBeauvoir2000a}. 
For each day of data collection, a single extrapolation is generated (an example is shown in Fig.~\ref{extrapolation and scatter}). 

The AC Stark effect introduces a shift in the resonance frequency of the $2$S$_{1/2}-8$D$_{5/2}$ transition that is linear with intensity \cite{Haas2006}. However, extrapolations on an ensemble of metastable atoms acquire a slight nonlinearity due to the different intensity profiles sampled by metastable atoms with different trajectories. As atoms following trajectories sampling the most intense portions of the 778 nm cavity mode begin to saturate, atoms along trajectories which sample lower intensities begin to contribute relatively more to the determination of the line center. This effect was present in the previous measurement of the $2$S$_{1/2}-8$D$_{5/2}$ transition~\cite{DeBeauvoir1997, DeBeauvoir2000a}, though the nonlinearity present in our extrapolations is smaller due to the more stringent geometric constraints on the atomic trajectories in our apparatus. 

While the nonlinearity is relatively small, ignoring it shifts the extrapolated resonance frequency by several kHz. Sampling an appreciable range of intracavity laser powers is required to properly determine the nonlinearity. From analytic considerations we have found that the nonlinearity acquired due to this spatial distribution is predominantly cubic and that this conclusion does not depend on the metastable spatial distribution (see SM); both of these results are strongly supported by our numeric lineshape models. Seventeen suitable extrapolations form the basis of our data set, and are shown in Fig.~\ref{extrapolation and scatter}.

Our second leading systematic is due to the DC Stark effect, which leads to shifts and distortions of the $2$S$_{1/2}-8$D$_{5/2}$ line. We have taken steps to passively mitigate stray fields by enclosing the entire metastable excitation and spectroscopic volume within a Faraday cage coated in colloidal graphite \cite{Porter1936}. Due to the near degeneracy of the 8D, 8P, and 8F manifolds, the transition is very sensitive to static fields with shifts of $\sim$12 kHz/(V/m)$^2$. Higher lying $n$-manifolds are even more sensitive to the presence of static fields due to the narrower natural linewidths, increasing degeneracy of the states, and larger dipole matrix elements between the states. This makes the line distortions of transitions to higher $n$-states a sensitive probe of the stray fields \cite{DeBeauvoir2000a}. 

\begin{figure} [t!]
    \includegraphics[width=.99\linewidth]{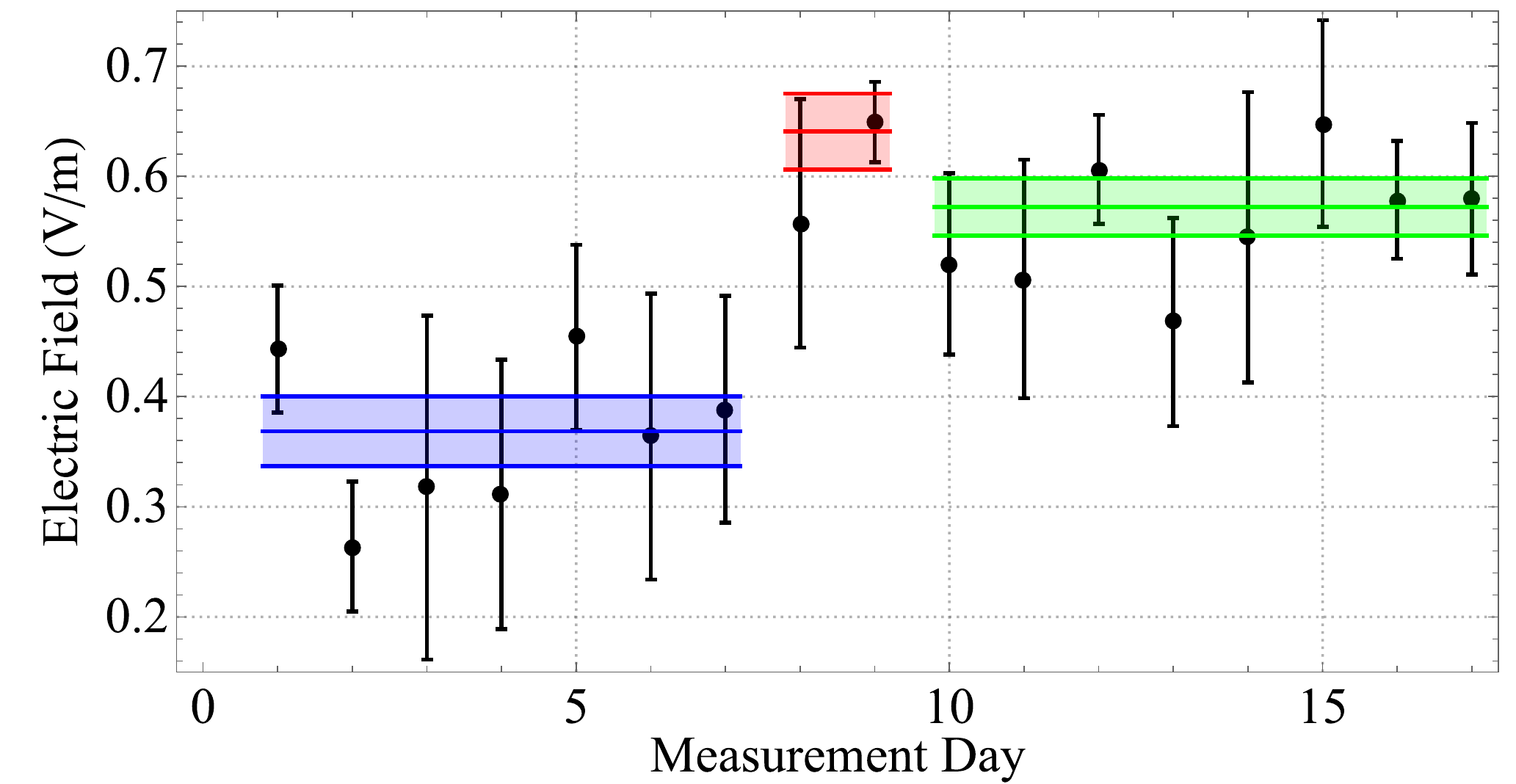}
    \caption{Fit electric field for each measurement day, with data collection batch average overlaid.}
    \label{Electric fields}
\end{figure}

From measurements of the $2$S$_{1/2}-12$D$_{5/2}$ lineshape distortion, we have observed that the stray fields are stable day-to-day as long as the system remains under vacuum, and that the stray field orientation is parallel with the atomic beam. The orientation of the stray field was determined by varying the excitation light polarization and comparing the shift and distortion of the line. Between batches of data collection on the $2$S$_{1/2}-8$D$_{5/2}$ line, the chamber was vented, leaving the possibility of stray field variation within the data set. Therefore, we also measured the electric field strength for each day \textit{in situ} by averaging several $2$S$_{1/2}-8$D$_{5/2}$ scans from a single day and at similar 778 nm laser power. We then fit this averaged line with the numeric model to match the subtle line distortion and extract the stray DC field (see SM for additional detail). Fig.~\ref{Electric fields} shows 
the determined electric field for each measurement day. We find the average static field for each batch of data and use that field strength to determine the shift for that set of data.

The coupling introduced by the DC electric field causes a quadratic shift due to nearby dipole-allowed transitions and a lineshape distortion due to the mixing between the nearly degenerate $8$D$_{5/2}$ and $8$F$_{5/2}$ states \cite{DeBeauvoir2000a}. Since the lines are fit with the simple analytic function given in Eqn.~(\ref{fit function}), the distortion also produces an additional shift. For each of the three measured electric fields shown in Fig.~\ref{Electric fields}, the resulting DC Stark shift is determined with the numeric model. Then, the appropriate correction for each extrapolated resonance frequency in a given batch is applied. The three batches of electric field corrections are -1.92(0.32) kHz, -5.45(0.54) kHz, and -4.43(0.37) kHz, respectively. Due to the statistical contribution of each batch to the final dataset, the weighted DC Stark correction for the entire measurement is -3.54(0.37) kHz.

The correction of -3.54(0.37) kHz assumes a stray field which is perpendicular to the spectroscopy light polarization. However, we cannot completely exclude a small parallel field, which can modify the required correction by up to 0.65 kHz. Additionally, from the numeric model, we find there is a small amount of cross-talk between the AC extrapolation and DC Stark effect. This cross-talk amounts to a -0.35 kHz correction to the DC Stark shift, and we assign the full 0.35 kHz shift as the associated uncertainty. Since the systematic shifts associated with the cross-talk effect, the possibility of non-perpendicular fields, and the statistical variance of the DC field correction are highly correlated, we combine their uncertainties linearly, and obtain a net DC Stark correction of -3.89(1.36) kHz. 

We believe the vacuum pressure is limited by water, which has an intrinsic dipole moment. During a collision with a water molecule, a hydrogen atom experiences a varying electric field which can drive population in the 8D state to nearby states, quenching the 8D state and broadening the line \cite{Sobelman1955}.  We have employed Monte Carlo simulations of these H-H$_2$O collisions to estimate the shift and broadening, similar to \cite{Matveev2019}. From these simulations we have found that any associated pressure shifts are below the $\sim$1 Hz level and insignificant at our current level of precision. 

Due to the two-photon excitation of the $2$S$_{1/2}-8$D$_{5/2}$ transition in an optical cavity, the first-order Doppler shift is effectively absent. The second-order Doppler shift remains, given by $\Delta\nu_{\text{DS}}=-0.5 (v/c)^2 \, \nu$, for atomic velocity $v$. An advantage of our apparatus is the ability to directly characterize the hydrogen and metastable hydrogen velocity distributions via a time-of-flight measurement \cite{Cooper2020}. From such measurements, we have found the metastable velocity distribution is well approximated by $P(v)\propto v^4e^{-\beta v^2}$, with $\beta=m/2k_bT$ and the required correction is $-0.73 (0.10)$ kHz.

\begin{figure}[b!]
    \includegraphics[width=0.99\linewidth]{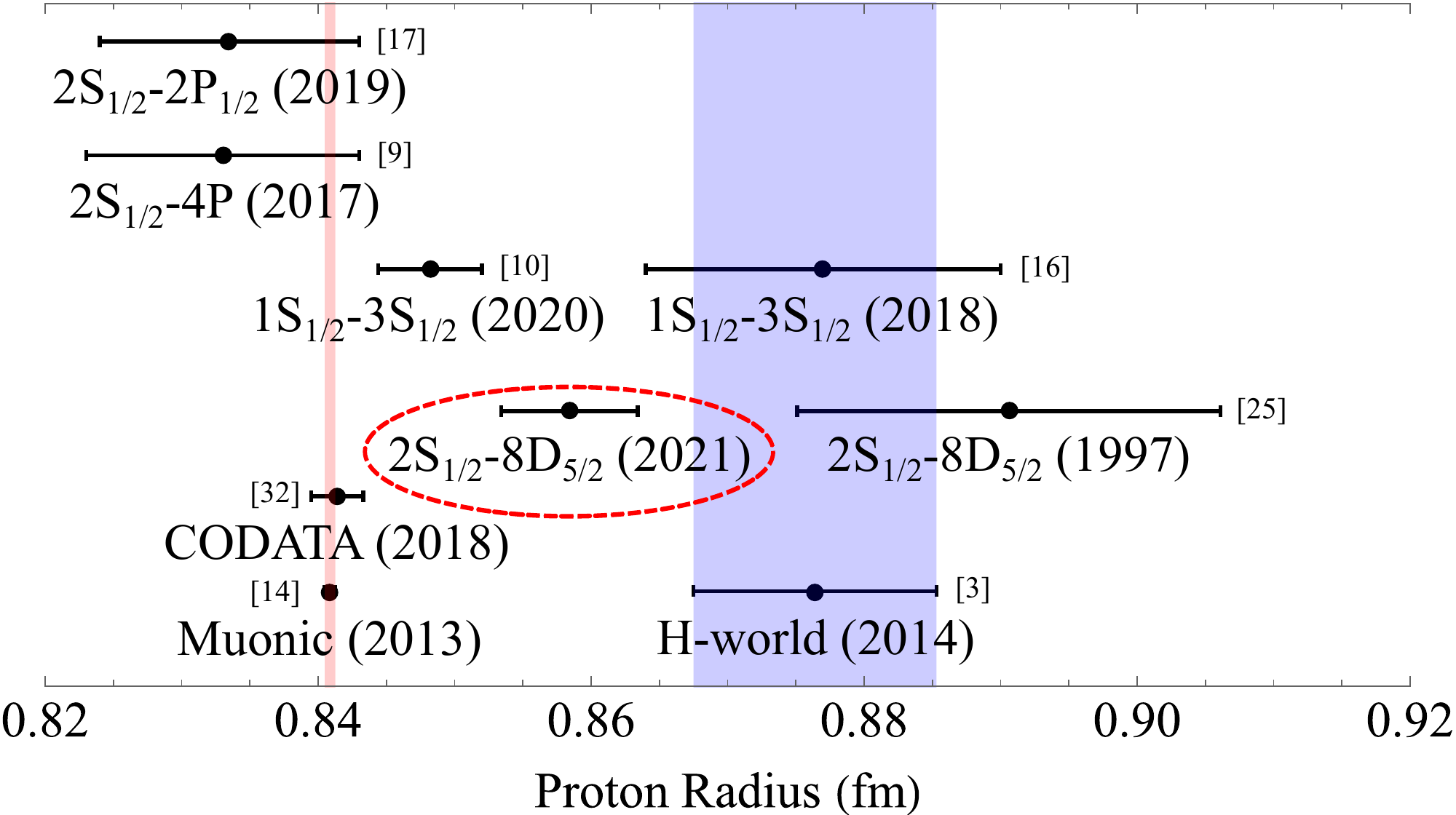}
    \caption{A selection of recent determinations of the proton radius when combining laser spectroscopy with the 1S$_{1/2}$-2S$_{1/2}$ transition \cite{Parthey2011, Beyer2017,Fleurbaey2018, Grinin2020, DeBeauvoir1997}. The result presented here is circled. The $r_p$ denoted by H-world (2014) corresponds to the proton radius obtained in \cite{Mohr2016} using only hydrogen spectroscopy data (Adj.~8 Table XXIX). Also shown are the proton radius determinations from Lamb shift measurements in hydrogen \cite{Bezginov2019} and muonic hydrogen \cite{Pohl2013}, along with the most recent CODATA value \cite{Tiesinga2021}.}
    \label{proton radius}
\end{figure}

From the AC extrapolation data, we recover a 2S$_{1/2}-$8D$_{5/2}$ hyperfine centroid of $770649561574.90(1.20)$ kHz. The DC Stark correction shifts this centroid by $-3.89(1.36)$ kHz to $770649561571.01(1.82)$ kHz. We then apply minor corrections in our uncertainty budget as shown in Table~\ref{budget}. A summary of the treatment of these corrections may be found in the SM. With all the systematic corrections accounted for, we find a $2$S$_{1/2}-8$D$_{5/2}$ resonance frequency of 
\begin{equation}
    \nu=770649561570.9(2.0)\hspace{2 pt} \text{kHz}.
\end{equation}
Combining our result with the $1$S$_{1/2}-2$S$_{1/2}$ value \cite{Parthey2011}, we obtain $r_p=0.8584(51)$ fm, and $R_\infty=10973731.568332(52)$ m$^{-1}$. Our obtained $r_p$ is presented alongside a selection of recent determinations of $r_p$ from spectroscopic results in Fig.~\ref{proton radius}. Our value is 3.1 combined standard deviations from the latest CODATA recommended $r_p$ value \cite{Tiesinga2021}.

\begin{table}[t!]
    \centering
    \caption{Minor corrections and uncertainties of the extrapolated 2S$_{1/2}$-8D$_{5/2}$ hyperfine centroid.}
    \begin{tabular}{lcc}
   &$\Delta\nu$ (kHz)&$\sigma$ (kHz)\\
    \hline
    \hline
        Stark Corrected \hspace{3 pt}& 770649561571.01&1.82\\
        \hline
        2$^{\text{nd}}$Order Doppler& 0.73 & 0.10 \\
        Zeeman Effect& 0& 0.56\\
        Frequency Calibration& -0.40& 0.47 \\ 
        Black-body Radiation& -0.49& 0.16\\
        Pressure Shifts& 0&$10^{-3}$\\
        8D$_{5/2}$ Hyperfine Structure & 0& 0.03 \\
        Photodiode Imperfections&0& 0.32 \\
        Incoherent Line Pulling& 0&$10^{-3}$\\
        Light Force Shift & $0$& $10^{-3}$\\
     \hline
        Total: Minor corrections & -0.16 & 0.82 \\
        \hline
        \hline
        Hyperfine Centroid& 770649561570.9&2.0\\
    \end{tabular}
\label{budget}
\end{table}

\begin{figure}[t!]
    \includegraphics[width=0.99\linewidth]{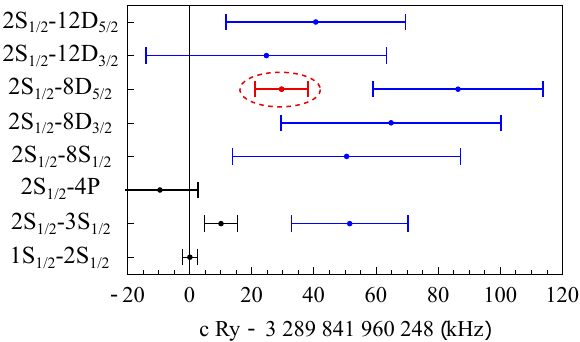}
    \caption{A comparison of the Rydberg constant determined through Eqn \ref{hydrogen energy states} from one experimentally determined hydrogen interval.  These determinations require a precise $r_p$ value, which was taken from the muonic hydrogen measurement alone \cite{Pohl2013}.
    Blue points were measured in Paris (2S-3S \cite{Fleurbaey2018},  2S-8S/D \cite{DeBeauvoir1997}, 2S-12D \cite{DeBeauvoir2000a}), black points were measured in Garching (1S$_{1/2}$-2S$_{1/2}$ \cite{Parthey2011}, 2S$_{1/2}$-3S$_{1/2}$ \cite{Grinin2020}, 2S$_{1/2}$-4P \cite{Beyer2017}), and our result is circled in red. As described in the text, we have subtracted the 1S$_{1/2}$-2S$_{1/2}$ interval \cite{Parthey2011} from the 1S$_{1/2}$-3S$_{1/2}$ measurements \cite{Grinin2020, Fleurbaey2018}}
    \label{Rydberg_Comparison}
\end{figure}

While the data shown in Fig.~\ref{proton radius} combines the $1$S$_{1/2}-2$S$_{1/2}$ measurement \cite{Parthey2011} with other hydrogen spectroscopy to extract $r_p$, it is also interesting to use the value of $r_p$ determined from muonic hydrogen \cite{Pohl2013} as input data for Eqn.~\ref{hydrogen energy states}.  This is a compelling approach considering that a recent measurement of the Lamb shift in normal hydrogen is in agreement with the muonic result \cite{Bezginov2019}. With that, a single measured interval in hydrogen is sufficient to extract $R_\infty$.  The result of this analysis for a selection of the most precise hydrogen laser spectroscopy data is shown in Fig. \ref{Rydberg_Comparison}. The uncertainty for the Rydberg constant determination from the  1S$_{1/2}$-2S$_{1/2}$ transition shown in Fig. \ref{Rydberg_Comparison} is almost entirely due to the theoretical uncertainty of the $1S$ state. In order to avoid including this correlated uncertainty in the 1S$_{1/2}$-3S$_{1/2}$ determinations \cite{Fleurbaey2018, Grinin2020}, we have subtracted the very precisely determined 1S$_{1/2}$-2S$_{1/2}$ transition frequency \cite{Parthey2011} to obtain 2S$_{1/2}$-3S$_{1/2}$ intervals. 

As can be seen from the Fig.~\ref{Rydberg_Comparison}, there is a general trend towards larger Rydberg constant when using experimentally determined intervals between states with larger $n$. It is interesting to note that hydrogen spectroscopy can provide a test for massive bosons that provide an additional coupling between the nucleus and electron \cite{Karshenboim2010,Jones2019}. Such bosons introduce a potential with finite range (a Yukawa potential) which affects certain $n$-states in hydrogen more strongly than others, producing an $n$-dependence when extracting $R_\infty$. As shown in the SM, the variation in Rydberg constant tends to decrease as the $n$ of either the upper or lower state increases.  We found that the reduced $\chi^2$ for the data shown in Fig. \ref{Rydberg_Comparison} decreases from $\sim$4.0 to $\sim$2.0 with the addition of a Yukawa potential with a length scale of $\sim$34 $a_0$.  Therefore, while the perturbation from such a potential can drastically reduce the inconsistency within the dataset shown in Fig.~\ref{Rydberg_Comparison}, it does not eliminate it.

Future investigations of $2$S$-n$S/D transitions in hydrogen are attractive due to the narrow natural lines afforded by such states and the convenience of the laser wavelengths required to drive the transitions. Additionally, the current $n$-dependence of the Rydberg constant extractions as shown in Fig. \ref{Rydberg_Comparison} provides a compelling case for further measurements of transitions to relatively high $n$ as a search for new physics \cite{Karshenboim2010, Jones2019} 

\begin{acknowledgments}
We gratefully acknowledge useful conversations with J. Berger, S. M. Brewer, A. Grinin, S. A. Lee, L. Maisenbacher, R. Pohl, J. L. Roberts, Th. Udem, and V. Wirthl.  We would like to thank the NIST WWVB team for accomodating our check of the GPS disciplined Rb time standard as described in the letter.  Finally we gratefully acknowledge funding for this measurement through the NIST Precision Measurement Grant (60NANB16D270), and the NSF CAREER Award (1654425).
\end{acknowledgments}

\end{document}



\title{Supplement for:\\
Measurement of the 2S$_{1/2}-$8D$_{5/2}$ transition in hydrogen}

\author{A. D. Brandt}

\author{S. F. Cooper}%
\author{C. Rasor}
\author{Z. Burkley}
 \altaffiliation[Now at ]{Dep. Physik, ETH Z\"{u}rich }
\author{D. C. Yost}
\email{Dylan.Yost@colostate.edu}

\affiliation{%
Department of Physics, Colorado State University, Fort Collins, CO 80523
}%

\author{A. Matveev}
 
\affiliation{Russian Quantum Center,  Skolkovo, Moscow 143025, Russia}%

\date{\today}

\maketitle

\section{Data Analysis}
\subsection{Lineshape fitting function}
In recent laser spectroscopy of hydrogen, experimentally obtained resonances have been fit with physically-motivated analytic functions \cite{Beyer2017, Grinin2020}, or numeric lineshape models \cite{Fleurbaey2018}. In this measurement, we have chosen the former.  

To obtain these analytic functions, we begin with the optical Bloch equations. For the 2S-8D transition, the 8D atoms preferentially decay back to the 1S state through $nP$ states, with only about 5\% decaying back to the 2S state \cite{DeBeauvoir2000a}. If we assume that the population in the upper state is small and that the coherences adiabatically follow the populations, the optical Bloch equations may be approximated as

\begin{equation}
    \dot\rho_{gg}=-\Omega \, \text{Im}(\rho_{ge})
    \label{densityeq1}
\end{equation}
\begin{equation}
    0=-i \, 2\pi  \Delta\nu \, \rho_{ge}+i\frac{\Omega}{2}\rho_{gg}-\frac{\gamma}{2}\rho_{ge}
\end{equation}
\begin{equation}
   0=\Omega \, \text{Im}(\rho_{ge})-\gamma\rho_{ee}
    \label{densityeq3},
\end{equation}
where $\Omega$ is the two-photon Rabi frequency, $\gamma$ is the decay rate from the excited state, and $\Delta\nu$ is the detuning from the two-photon resonance. Since we are detecting the population remaining in the ground state, we decouple the differential equations to find 
\begin{equation}
    \dot\rho_{gg}=-\frac{1}{4}\frac{\gamma\hspace{1 pt}\Omega^2}{(2\pi\Delta\nu)^2+(\gamma/2)^2}\rho_{gg}.
\end{equation}
The solution, assuming a time-dependent Rabi frequency, is given by
\begin{equation}
    \rho_{gg}(t)=\text{exp}\left(-\frac{1}{4}\frac{\gamma\hspace{1 pt}\int_{0}^{t} \Omega^2 d\tau}{(2\pi\Delta\nu)^2+(\gamma/2)^2}\right)
    \label{fit function approx 1}.
\end{equation}
Here, we have assumed that the detuning is constant, which is only approximately true due to the varying AC Stark shift as the atom traverses the laser beam.
From Eqn.~\ref{fit function approx 1}, we find that a physically motivated fit function is given by
\begin{equation}
    \mathcal{F}(A,\alpha,\nu_0,\gamma_0)=A \, \text{exp}\left[- \mathcal{L}_0(\alpha,\nu_0,\gamma_0)\right],
\end{equation}
where $\mathcal{L}_0$ is a Lorentzian function with width $\gamma_0$, $\nu_0$ is the center frequency of the transition, $\alpha$ is the Lorentzian amplitude, and $A$ accounts for the off-resonant metastable count rate.

The 2S$_{1/2}-$8D$_{5/2}$ transition also contains unresolved hyperfine structure that must be taken into account (F=2 and F=3). To account for this structure, we adjust our fitting function to give
\begin{equation}
    \mathcal{F}(A,\alpha_2,\alpha_3,\nu_2,\nu_3,\gamma_2,\gamma_3)=A \, \text{exp}\left[-  \mathcal{L}_0(\alpha_2, \nu_2,\gamma_2) - \mathcal{L}_0(\alpha_3,\nu_3,\gamma_3)\right],
    \label{fit_function}
\end{equation}
where the subscript denotes the hyperfine component of the 8D$_{5/2}$ level. The ratio of $\alpha_2$ and $\alpha_3$ is given by $\alpha_3/\alpha_2=3.5$ due to the relative strengths of the transitions -- this is independent of laser polarization.  In addition, the natural linewidth and transit-time broadening are both independent of the hyperfine state so that we can safely assume that $\gamma_2=\gamma_3$. Finally, the hyperfine splitting of the 8D$_{5/2}$ state is known and given by $\text{HFS}=142.43$ kHz \cite{Kramida2010} which gives a fixed relationship between $\nu_3$ and $\nu_2$ -- in practice we use the detuning from the centroid frequency, $\nu_c$, as the fit parameter. Therefore, we fit our experimental lines with
\begin{equation} 
    \mathcal{F}(A, \alpha, \nu_c, \gamma)= A \, \text{exp}[-\alpha(\mathcal{L}_2(\nu_c,\gamma)+\mathcal{L}_3(\nu_c,\gamma))],
    \label{fit function}
\end{equation}
where the ratio of the amplitudes of the Lorentzian functions ($\mathcal{L}_2$ and $\mathcal{L}_3$) are in a fixed ratio of 1 to 3.5. An example of a single scan fit with Eqn.~\ref{fit function} is shown in supplemental Fig.~\ref{line fitting}. 

\begin{figure} [h!]
    \centering
    \includegraphics[width=.6\linewidth]{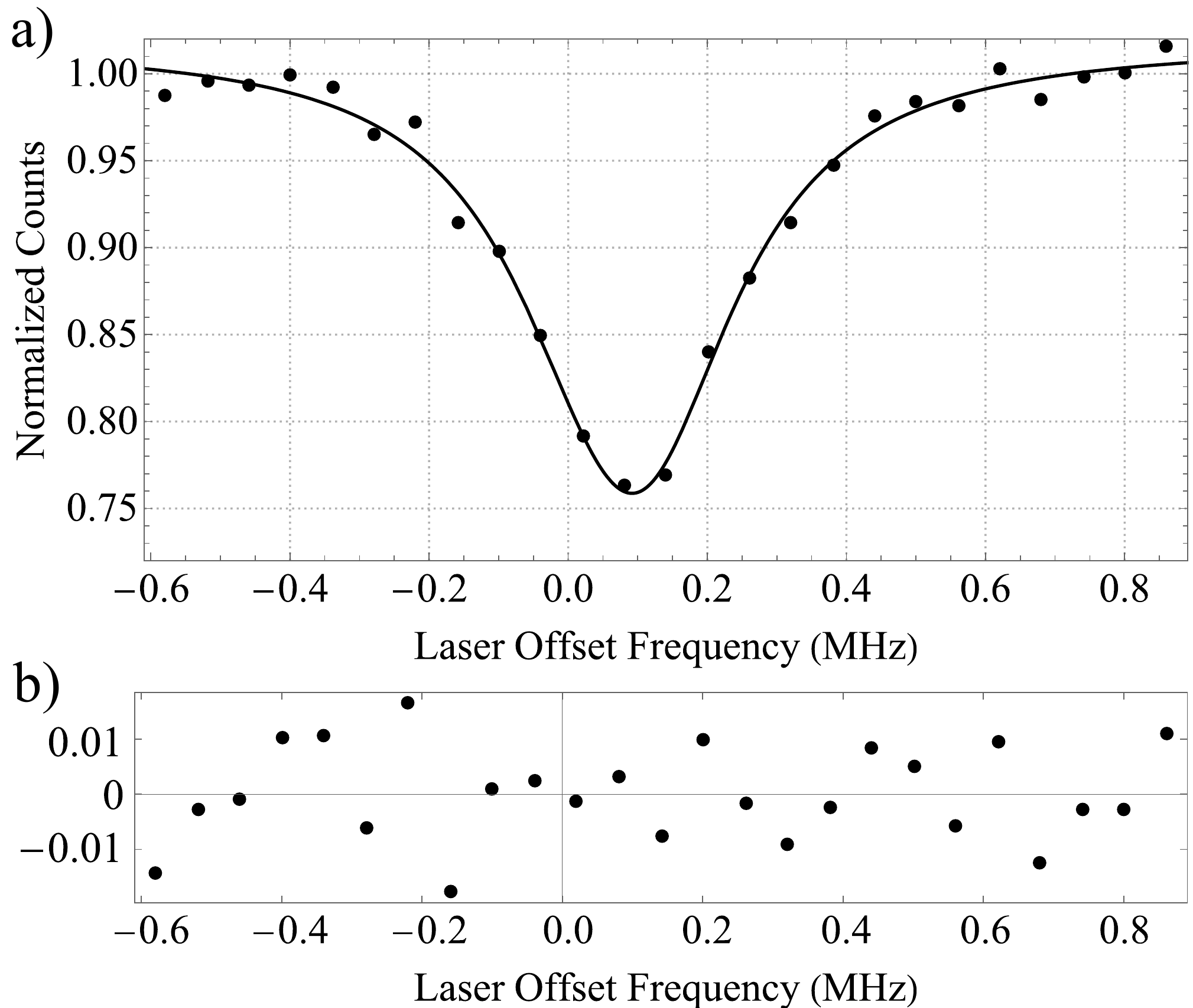}
    \caption{A single scan of the $2$S$_{1/2}-8$D$_{5/2}$ line fit with Eq.~\ref{fit function}. The off-resonant count rate normalized to unity. A typical off-resonant count rate is $\approx 10^4$/s. a) A single scan fit and b) a single scan fit residuals. A shot noise limited signal-to-noise ratio of $\approx 10^2$ is expected for the given count rates. }
    \label{line fitting}
\end{figure}

\subsection{The AC Stark shift extrapolations}
To account for the AC Stark shift, we perform scans of the resonance at various laser powers, determine the centroid frequencies for individual scans, and extrapolate these to zero laser power.  For these extrapolations, we again use simple, analytic functions, which we justify below.

The shift due to the AC Stark effect can be written as
\begin{equation}
    \Delta\nu_{ac}=\beta_{ac}I(t)
\end{equation}
where $\beta_{ac}$ is the AC Stark coefficient \cite{Haas2006}.
While the AC Stark shift is itself linear, the extrapolations can exhibit nonlinear behavior when measuring ensembles of trajectories and when appreciable saturation of the transition occurs. In the case of many atomic trajectories contributing to the overall line, the contribution of each trajectory is weighted by the population driven in that particular trajectory. Therefore, the average AC Stark shift is
\begin{equation}
    \langle\Delta\nu_{ac}\rangle=\frac{\int \mathcal{D}(s)\,(1-\rho_{gg}(s))\Delta\nu_{ac}(s)ds}{\int\mathcal{D}(s)\, (1-\rho_{gg}(s))ds},
    \label{weightacshift}
\end{equation}
where $\mathcal{D}(s)$ represents the spatial distribution of atomic trajectories, and $s$ is the transverse distance from the peak of the 778 nm Gaussian intensity profile. The spectroscopy signal is determined by counting the quenched fraction of the metastable atoms, given by $1-\rho_{gg}(s)$. From the analysis of the 2S$_{1/2}-$8D$_{5/2}$ lineshapes above, we find a functional form for the spectroscopy signal given by
\begin{equation}
    1-\rho_{gg}(s)=1-\text{exp}[-\kappa P_0^2 e^{-k s^2}],
    \label{signal}
\end{equation}
where $\kappa$ and $k$ are scaling constants, and $P_0$ is laser power within the optical cavity.  This follows from Eq.~(\ref{fit function approx 1}) and the fact that $\Omega$ is proportional to $P_0$ for two-photon transitions \cite{Haas2006}. For low intracavity power, $\rho_{gg}(s)\approx 1-\kappa e^{-k s^2}P_0^2$ and
\begin{equation}
     \langle\Delta\nu_{ac}\rangle=\frac{\int \mathcal{D}(s)  e^{-ks^2}P_0^2 (P_0\beta_{ac}e^{-ks^2})ds}{\int \mathcal{D}(s) e^{-ks^2}P_0^2 ds}=P_0\frac{\int f(s)ds}{\int g(s)ds},
\end{equation}
where $f(s)$ and $g(s)$ are spatially dependent functions without $P_0$ dependence.  From this we see that $\Delta\nu_{ac}$ is proportional to the intracavity laser power and a linear extrapolation is suitable. 

We may estimate the effect of saturation by keeping an additional term in the series expansion of $1-\rho_{gg}(s)$. In that case, $(1-\rho_{gg}(s))\approx \kappa e^{-ks^2}P_0^2 -\frac{1}{2}\kappa^2 \, e^{-2ks^2}P_0^4$. Substituting this approximation of $1-\rho_{gg}$, we find 
\begin{equation}
     \langle\Delta\nu_{ac}\rangle \approx h_1 \, P_0 - h_2 \, P_0^3,
     \label{cubic_function}
\end{equation}
where $h_1$ and $h_2$ are constants determined by the spread of metastable atomic trajectories and the width of the spectroscopy laser beam's transverse Gaussian profile. Therefore, the nonlinearity introduced to the AC extrapolation is predominantly cubic. One should note that this conclusion does not rely on the exact form of the metastable distribution, $\mathcal{D}(s)$, and the accuracy of this result has been confirmed with the numeric model (described in the next section).  

\subsection{Numeric Model}

The numeric model is used primarily for quantifying the DC Stark shifts and to verify that fitting functions described above are sufficiently accurate. We have developed a numeric model based on density matrices, which allows for a description of repopulation of the 2S state, and a model based on state amplitudes. We have found that the repopulation of the 2S state via spontaneous emission produces shifts of $<\pm10$ Hz in the extrapolated AC-Stark-shift-free resonance frequency. Therefore, we focus on the state amplitude treatment here, since that treatment allows for the inclusion of DC Stark effects in a much simpler manner.

Roughly 15 cm before the spectroscopic region, the 2S state is populated via a two-photon excitation with 243 nm laser radiation tuned to the 1S$_{1/2}^{F=1}$ to 2S$_{1/2}^{F=1}$ transition.  The hyperfine splitting of the 1S and 2S states is large enough that there is effectively no population in the $2S^{F=0}$ state. We assume that our ground state (1S) atoms are equally populated in the $F=1$, $m_F=0, \pm1$ levels. Since, the 1S$^{F=1}_{m_F=\pm1}\rightarrow$2S$^{F=1}_{m_F=\pm1}$ and 1S$^{F=1}_{m_F=0}\rightarrow$2S$^{F=1}_{m_F=0}$ matrix elements have the same magnitude \cite{Haas2006}, the 2S $F=1$ hyperfine manifold is equally populated as well.  The lifetime of the 2S$_{1/2}$ state is $\approx 122$ ms so no appreciable decay occurs in the $\approx 1$ ms  transit time of the metastable beam through the apparatus. We choose to drive the 2S$_{1/2}$-8D$_{5/2}$ transition with linearly polarized light, which defines the $z$-axis in the experiment. The two-photon selection rules for $\Delta L=2$ using linearly polarized light, and with the laser polarization defining quantization axis, dictate that $\Delta m_F=0$ \cite{Haas2006}. 

The couplings introduced by the DC electric field causes a quadratic shift due to nearby dipole-allowed transitions and a lineshape distortion due to the mixing of the nearly degenerate 8D$_{5/2}$ and 8F$_{5/2}$ states \cite{DeBeauvoir2000a}. The quadratic shift due to nearby dipole-allowed transitions is calculated with second-order perturbation theory. Both the mixing and the quadratic shift are included in the state amplitude equations.  

\begin{figure}
    \centering
    \includegraphics[width=.95 \linewidth]{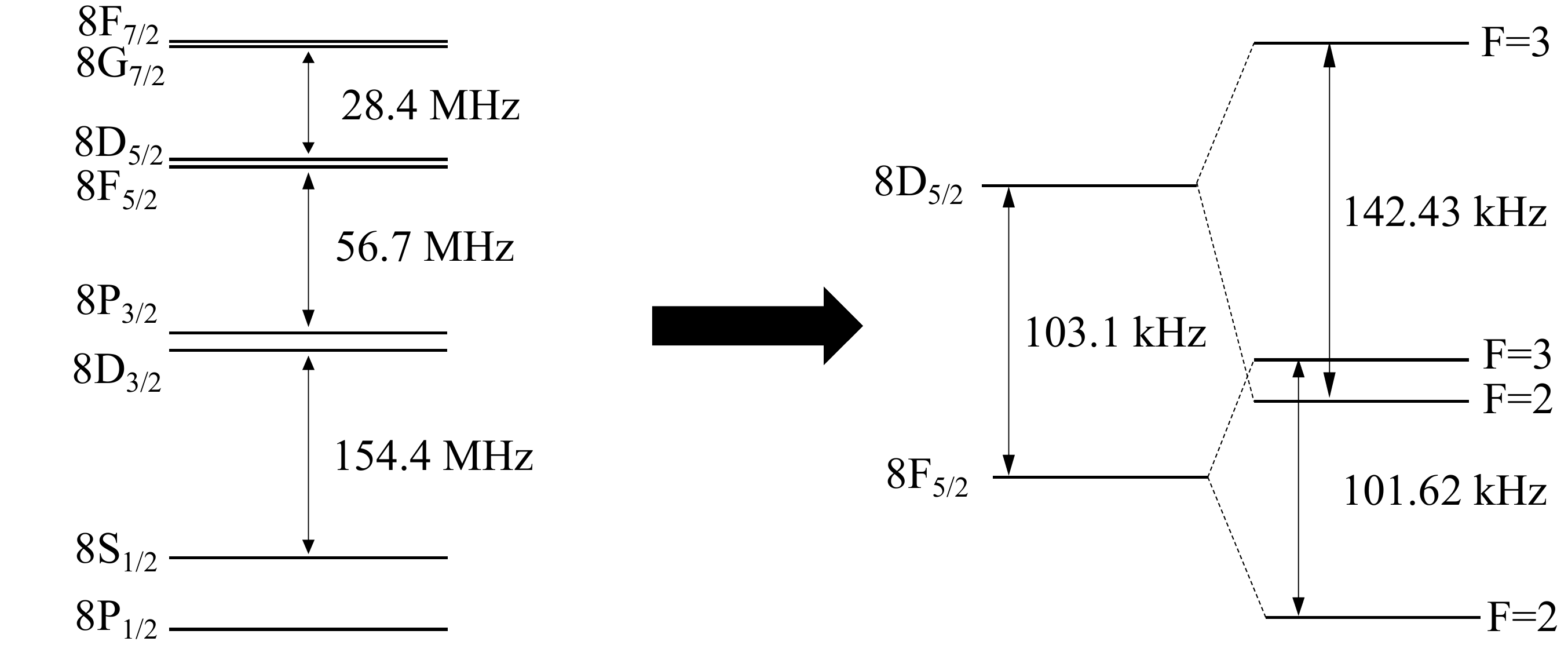}
    \caption{Level structure of the relevant states in the $n=8$ manifold.}
    \label{n8 structure}
\end{figure}

We label the state amplitudes for the hyperfine components of the 2S$_{1/2}$, 8D$_{5/2}$, and 8F$_{5/2}$ levels with $g$, $e$, and $f$ respectively. The coupled set of differential equations describing the interaction is given by

\begin{equation}
    \dot{c}_g=-i\sum_e \frac{\Omega_{ge}}{2}c_e
    \label{amplitude + stark 1},
\end{equation}

\begin{equation}
    \dot{c}_e=(i\Delta\omega_e-\frac{\gamma_e}{2})c_e-i\frac{\Omega_{ge}}{2}c_g-\frac{i}{\hbar}\sum_fU_{ef}c_f,
\end{equation}

\begin{equation}
    \dot{c}_f=(i\Delta\omega_f-\frac{\gamma_f}{2})c_f-\frac{i}{\hbar}\sum_eU_{fe}c_e,
    \label{amplitude + stark 3}
\end{equation}

\begin{equation}
    \Delta\omega_e=2\omega-(\omega_{ge}+\delta\omega^{(e)}_{\mathcal{E}}\mathcal{E}^2+\delta_{ge}(t)),
\end{equation}

\begin{equation}
 \Delta\omega_f=2\omega-(\omega_{gf}+\delta\omega^{(f)}_{\mathcal{E}}\mathcal{E}^2+\delta_{gf}(t)),
\end{equation}
where $\omega$ is the laser frequency, $\omega_{ij}=\omega_j-\omega_i$ is the resonance frequency separating states $\ket{i}$ and $\ket{j}$, $\mathcal{E}$ is the magnitude of the stray electric field, $\Omega_{ij}$ is the trajectory-dependent Rabi frequency \cite{Haas2006}, $\delta\omega^{(i)}_{\mathcal{E}}$ is the quadratic Stark shift coefficient for state $\ket{i}$, $\delta_{ij}$ is the combined AC Stark and second order-Doppler detuning terms, and $U_{ij}$ is the perturbing matrix element connecting states $\ket{i}$ and $\ket{j}$ (i.e. $-\bra{i}\vec{\mathcal{E}}\cdot q\vec{r}\ket{j}$). The combined detuning of the AC Stark effect and the second-order Doppler effect is given by
\begin{equation}
    \delta_{ij}=2\pi[\beta_{ac}(j)-\beta_{ac}(i)]I(t,v,s)-\frac{v^2}{2c^2}\omega_{ij}, 
    \label{population detuning terms}
\end{equation}
where $\beta_{ac}(j)$ is the state dependent AC Stark shift coefficient. $I(t,v,s)$ is the time-dependent intensity seen by an atom traversing through the laser beam a distance $s$ away from the Gaussian beam center at velocity $v$, which is given by
\begin{equation}
I(t,v,s)=I_0\text{Exp}\left(-2\frac{v^2t^2 \text{sin}^2(\theta)}{w^2}-2 \frac{s^2}{w^2}\right).
\end{equation} 
The time-dependent Rabi frequency seen by an atom in a specific trajectory is likewise given by
\begin{equation}
    \Omega_{ij}(t)=2(2\pi\beta_{ij}) I(t,v,s),
    \label{2photonrabi}
\end{equation}
where $\beta_{ij}$ is the two-photon matrix element connecting states $\ket{i}$ and $\ket{j}$ \cite{Haas2006}. For each trajectory, we numerically integrate Eqns.~(\ref{amplitude + stark 1}-\ref{amplitude + stark 3}) to find a lineshape $\tilde{L}(\omega,s,v)$. To include the effect of the different trajectories, we integrate $\tilde{L}(\omega,s,v)$ over a  spatial and velocity distribution $\rho(s,v)$ to find the full linshape, given by
\begin{equation}
       L(\omega)=\iint\tilde{L}(\omega, s,v)\, \rho (s,v) \, ds \, dv.
\end{equation}
The function $\rho(s,v)$ is determined by time-of-flight measurements \cite{Cooper2020}, and the geometric constraints set by the metastable hydrogen/243 nm light overlap and the divergence of the atomic hydrogen beam.

The distortion present on the 2S$_{1/2}$-8D$_{5/2}$ and the 2S$_{1/2}$-12D$_{5/2}$ lineshapes is used to quantify the magnitude of the electric field. A single scan of the 2S$_{1/2}$-8D$_{5/2}$ resonance is insufficient to quantify the distortion, therefore many scans are averaged together before being fit with the numeric model. When averaging lines, we choose several lines with nominally equal photodiode voltage. For averaging, we have chosen a PD voltage that corresponds to quenching approximately 20\% of the total metastable flux.  We find that the electric fields determined \textit{in situ} on the 2S$_{1/2}$-8D$_{5/2}$ lines are consistent with the determinations of the electric field on the 2S$_{1/2}$-12D$_{5/2}$ lines. Examples of the 2S$_{1/2}$-8D$_{5/2}$ and 2S$_{1/2}$-12D$_{5/2}$ lines fit with the numeric model are shown in supplemental Fig.~\ref{E field characterization}.

\begin{figure} [h]
    \centering
    \includegraphics[width=.99\linewidth]{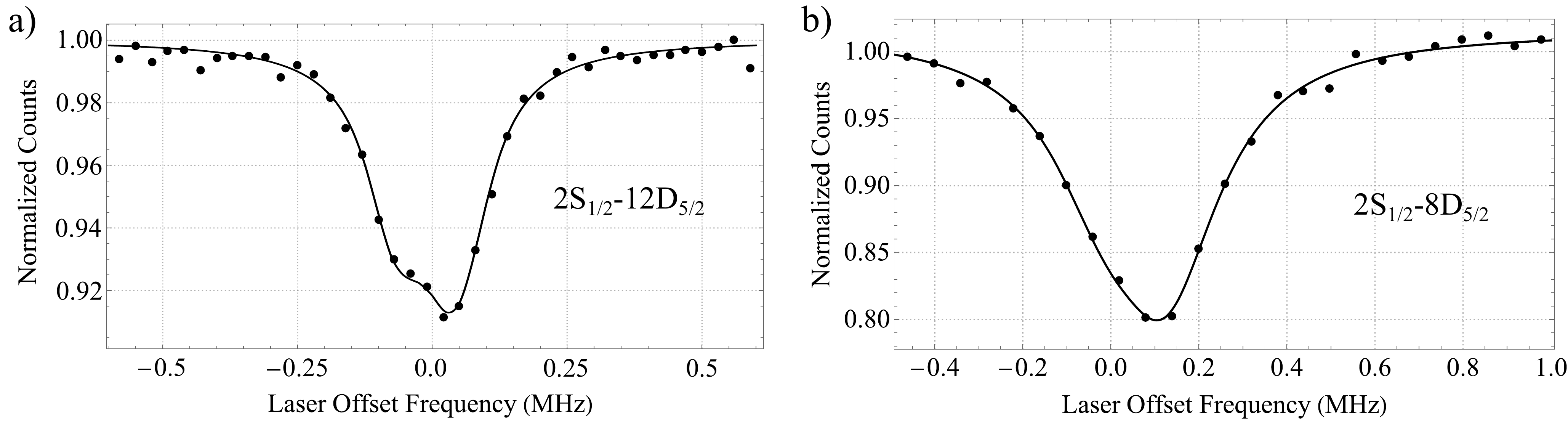}
    \caption{Electric field characterization. a) Example of measured $2$S$_{1/2}-12$D$_{5/2}$ lineshape fit with the numeric model. This corresponds to a field strength of 3 mV/cm b) Example of measured $2$S$_{1/2}-8$D$_{5/2}$ lineshape, composed of 27 scans averaged together fit with the numeric model. This line fit corresponds to a 6.5 mV/cm field, which is the largest field present in the data. }
    \label{E field characterization}
\end{figure}

\begin{figure} [h]
    \centering
    \includegraphics[width=.99\linewidth]{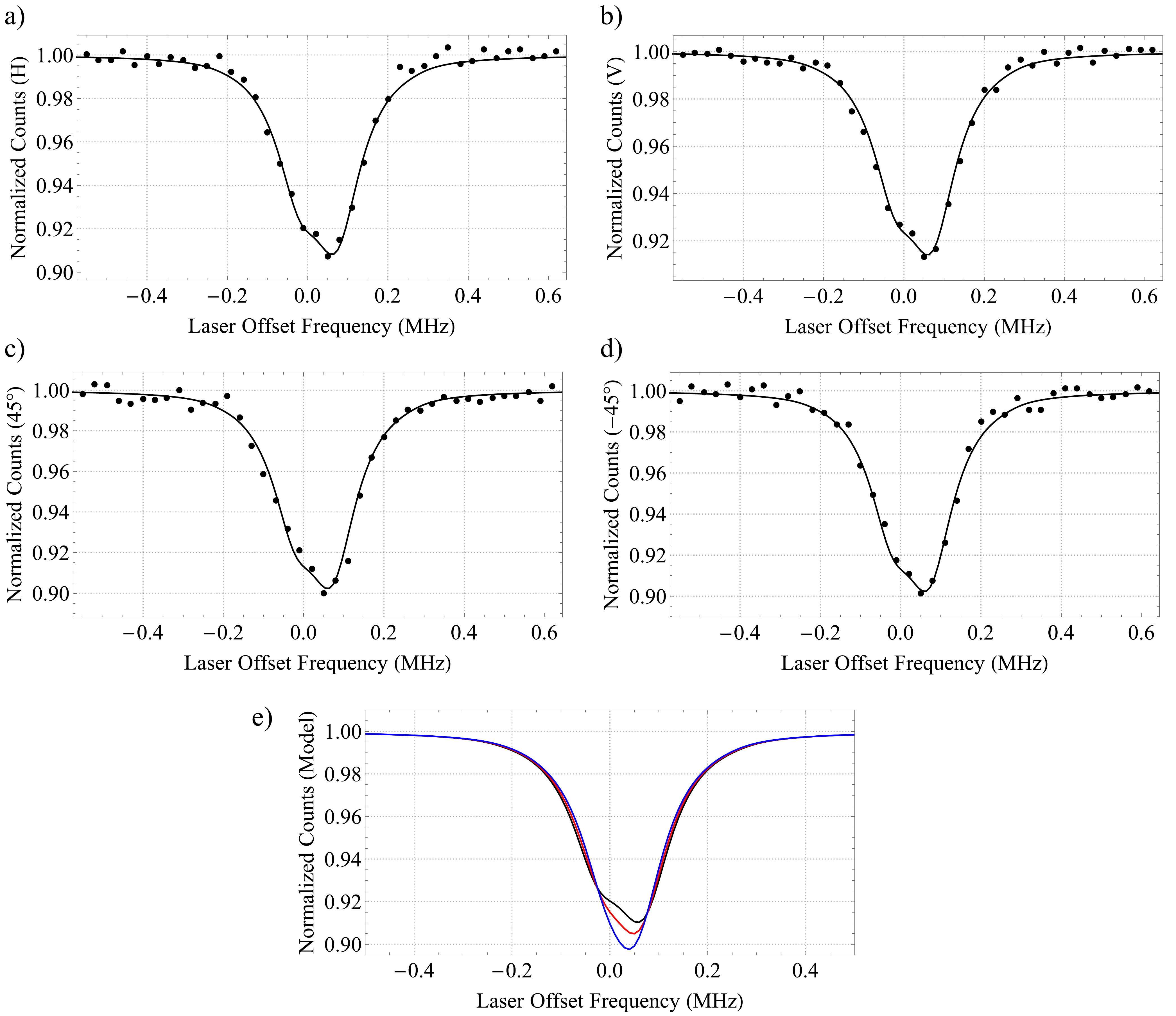}
    \caption{Example 2S$_{1/2}$-12D$_{5/2}$ lineshapes excited by distinct linear polarizations. We overlay the data with simulated data for a stray field of 2.5 mV/cm perpendicular to the polarization of the light field.  Each lineshape is composed of $\approx$30 averaged scans. a) Horizontal polarization. b) Vertical polarization. c) 45$^{\circ}$. d) -45$^{\circ}$. e) Numerically modeled 2S$_{1/2}$-12D$_{5/2}$ lines under the influence of a uniform field of 2.5 mV/cm. The relative angle between the light polarization and stray field is 90$^\circ$ (black), 55$^\circ$ (red), and 35$^\circ$ (blue).}
    \label{E field orientation}
\end{figure}

We find that the distortion present on the experimentally measured 2S$_{1/2}$-12D$_{5/2}$ lineshapes is insensitive to the polarization of the spectroscopy light. However, from investigations with the lineshape model, we find that the nature of the distortion is highly contingent upon the relative orientation of the stray field and the spectroscopy light. From this, we conclude that the stray electric field must be nearly perpendicular to all possible laser polarizations -- i.e. along the direction of the light propagation. In Fig.~\ref{E field orientation}, we show the 2S$_{1/2}$-12D$_{5/2}$ line excited by four linear polarizations (horizontal, vertical, $\pm$45$^{\circ}$) on a given day with the same lineshape generated by the numeric model overlaid. As shown in the figure, the distortion is consistent as the polarization is rotated. We have additionally included an example of the 2S$_{1/2}$-12D$_{5/2}$ modeled lineshape under the influence of a uniform electric field as the relative orientation of the light polarization and stray field is varied, which shows a clear variation in the lineshape as the relative angle changes. We have investigated this polarization dependence of the 2S$_{1/2}$-12D$_{5/2}$ line after multiple batches of data acquisition and have found similar insensitivity to the light polarization, indicating that the stray field orientation was nearly perpendicular to the light field throughout the experiment.  

\subsection{Statistics}
Before fitting each scan of the 2S$_{1/2}-$8D$_{5/2}$ resonance with the fit function, $\mathcal{F}(A, \alpha, \nu_c, \gamma)$, we first remove data points where there was a problem with excitation of the 1S-2S transition due to laser lock breaks, which manifest as far statistical outliers in the metastable count rate. Similarly, spurious lock breaks of the 778 nm power enhancement cavity are removed, which are evident on the 778 nm transmission diode. A scan of the 2S$_{1/2}-$8D$_{5/2}$ resonance is composed of 25 frequency points if no outliers are removed -- scans composed of fewer than 20 points are omitted entirely from the data analysis. In a typical day of data acquisition, less than $5\%$ of all individual data points are omitted. When fitting scans with $\mathcal{F}(A, \alpha, \nu_c, \gamma)$, each data point in the scan is weighted by the estimated shot noise for that point. 

The reduced chi-squared statistic is used to quantify to goodness-of-fit when analyzing individual resonance scans. We first assign error bars to the individual data points based on shot noise.  This results in a $\chi^2_k\approx 1.5$ on average.  Here, $k$ denotes the degrees of freedom, which is typically 21 for a full scan of 25 points with 4 fit parameters. While $\chi^2_k$ is typically greater than one, this is expected due to the contribution of some technical noise arising from variations in the atomic beam flux. Therefore, the error bars on the individual data points are scaled until $\chi^2_k=1$ for a given scan --  fluctuation of the metastable flux is not associated with any shift (possible density effects are discussed later and shown to be negligible). 

Each line fit results in a hyperfine centroid ($\nu_c\pm\sigma_c$) for a given 778 nm photodiode voltage $V$, which is proportional to the intracavity power. In a day of data acquisition, the set of $\{V^{(i)},\nu_c^{(i)}\}$ obtained from the scans of the resonance allow for an AC Stark extrapolation to zero intracavity power. The data set is fit with a cubic function of the form $\nu_c(V)=\nu_{0}+aV+bV^3$, with each ordered pair in the data set weighted by the associated $1/\sigma_c^2$ for that data point.  For these fits, we typically find $\chi^2_N\approx0.95$ for a given day. Since $\chi^2_N$ itself has a variance of $\sqrt{2/N}$, this is reasonable (a typical extrapolation has $N\approx 200$). For a given day's worth of data, we then correct the extrapolated centroid frequency for the DC Stark shift (see main text for further discussion). 

We obtained 17 days' worth of data (shown in the main text) and calculate the statistical mean of the resonance frequency, $\bar{\nu}$, and variance, $\bar{\sigma}_{\nu}$, for the entire set.  The variance of the mean frequency is given by $\bar{\sigma}_{\nu}=\left(\sqrt{\sum1/\sigma_{\nu,i}^2}\right)^{-1}=1.2$ kHz. The scatter within the full set is in excellent agreement with statistical fluctuations as we find a reduced chi-squared statistic of 1.05.  Therefore, there is no evidence of significant and varying systematic effects in the data set.

\section{Minor Systematic Corrections}

Our method for accounting for our two largest systematic effects, the AC and DC Stark shifts, is addressed in the main text and the preceding sections.  Here, we discuss smaller systematic effects.

\subsection{Second-order Doppler shift}
 
The second-order Doppler effect enters as $\Delta\nu_{DS}=-\frac{1}{2}(v/c)^2\nu$, where $v$ is the atom velocity in the lab frame, $c$ is the speed of light, and $\nu$ is the rest-frame resonance frequency. We have characterized the velocity distribution of our metastable hydrogem beam through time-of-flight measurements \cite{Cooper2020}. From these investigations, we have found that the velocity distribution is modified from a Maxwell-Boltzmann distribution by the excitation and detection dynamics to $P(v)\propto v^4 e^{-\beta v^2}$, with $\beta=m/2k_bT$. The temperature of the atomic beam is estimated with a silicon diode temperature probe mounted on the surface of the cryogenic nozzle. 

Investigations with the numeric lineshape model indicate that the net second-order Doppler shift on an ensemble of atoms is well-approximated by calculating the shift associated with the most-probable velocity class. The uncertainty associated with the most-probable velocity for a given nozzle temperature reading is $<10$ m/s. Fitting the time-of-flight measurements obtained in \cite{Cooper2020} gives a second-order Doppler shift of $-1475(25)$ Hz at 9.5 K and $-930(11)$ Hz at 5.9 K. Spectroscopy was performed on the atomic beam for temperatures between 4.5 K and 4.9 K, which corresponds to a second-order Doppler shift of -$0.73(10)$ kHz.  This shift is corrected for in our final measurement as indicated in the main text.

\subsection{Zeeman effect}
We passively mitigate Earth's magnetic field with a pair of concentric magnetic shields composed of high-permeability metal (mu-metal) centered around the 778 nm light/atom interaction region. We have simulated the attenuation of Earth's field in the center of the shields and we expect residual magnetic fields of less than 1 mG within the spectroscopic volume.  

The linear Zeeman effect causes a shift of a magnetic sublevel with quantum number $m_F$, which is given by
\begin{equation}
    \Delta E_Z/\hbar=-\bra{\psi}\vec{\mu}\cdot\vec{B}\ket{\psi}= \hspace{1pt} \frac{1}{\hbar}g_F\mu_Bm_F B,
    \label{linearzeeman}
\end{equation}
where $B$ is the magnetic flux density, $g_F$ is the g-factor, and $\mu_B$ is the Bohr magneton. With linearly polarized spectroscopy light the linear Zeeman effect only causes a broadening  of the resonance. However, with circularly polarized light oriented about a stray magnetic field, a shift can occur. 

In order to quantify the possibility of a residual Zeeman shift, we have also performed spectroscopy of the 2S$_{1/2}-$8D$_{5/2}$ transition with $\sigma^+$ and $\sigma^-$ light of similar intensity. The splitting of these two lines amounts to maximum of 3.7 kHz (limited by the statistics of the measurement), indicating that the maximal possible Zeeman shift is 1.9 kHz. From the maximal possible shift of 1.9 kHz, we estimate a magnetic field strength of 0.6 mG, which is in good agreement with our magnetic shields simulation.  

This maximal 1.9 kHz shift is suppressed by the polarization purity of the spectroscopy light. We have measured the polarization purity by determining the maximum polarization extinction ratio, given as $R$, with a polarizing beamsplitter after the 778 nm enhancement cavity and found that $R\leq0.01$. This leads to a maximum imbalance in the right versus left components of  $\approx$20\%.  By assuming this maximum imbalance in the transition amplitudes between all hyperfine components, we find that the measured 1.9 kHz shift should be suppressed by a factor of $\approx 3.4$, leading to a maximal shift of 0.56 kHz from the Zeeman effect. 

Since this represents an upper bound, we apply no shift and assume $\pm0.56$ kHz as the full uncertainty. Higher-order Zeeman corrections are negligible at our current level of precision.

\subsection{Frequency calibration and laser spectrum}
A GPS-trained Rb-oscillator (Stanford Research systems FS740) provides our absolute frequency reference.
This source is used to reference two frequency counters which redundantly monitor the frequency comb repetition rate. Our optical frequency comb is based on an erbium-fiber oscillator centered at 1550 nm. The comb offset frequency, $f_0$, is stabilized and a comb mode near 1550 nm is stabilized indirectly to an ultra-stable cavity with the help of a transfer laser oscillator. By measuring the repetition rate and phase locking $f_0$ of the comb, the absolute frequency of all comb teeth are determined. Since the Coherent-899 Ti:Sapphire and the 972 nm ECDL (quadrupled to 243 nm to excite the 1S-2S transition) are coherently phase-locked to the comb, we can determine the absolute frequency of those lasers from the RF beat note frequencies. 

The RF beat notes with the optical frequency comb can be converted into absolute atomic frequencies via
\begin{equation}
    \nu_{1S-2S}= 8(n f_r+f_0+f_{\text{beat}}^{972})
\end{equation}
for the 972 nm laser system and
\begin{equation}
    \nu_{2S-8D}=2(m f_r + 2f_0+f_{\text{beat}}^{778})
\end{equation}
for the 778 nm spectroscopy laser. The factor of two multiplying $f_0$ arises because we frequency double the frequency comb output before performing the beat note with the 778 nm light.  The repetition rate of the frequency comb is determined by the frequency counters over 100 s gates, and a typical day's data amounts to 5000-20000 s of $f_r$ counting.

In principle, an asymmetric spectral distribution of the spectroscopy laser could pull the 2S-8D resonance. The phase noise of the 778 nm light is strongly suppressed by phase-locking to the frequency comb. Nevertheless, we have quantified the phase noise distribution on the 778 nm light by investigating the beat note. By convolving the observed spectral density with our fit function, we estimate the laser noise amounts to a 5 Hz shift and 2 kHz of line broadening. Both are negligible effects.  

As mentioned in the main text, we have verified the absolute frequency calibration of the GPS-disciplined FS740 by locally counting a 5 MHz signal provided by the NIST WWVB station in Fort Collins, Colorado for 17 hours. We have found a fractional frequency counter offset of $+5\times10^{-13}$, which is within the expected performance of GPS-disciplined Rb-oscillators \cite{Lombardi2008}. A typical day's worth of data is composed of $10^4$ s of data, which corresponds to a frequency stability of $3\times10^{-13}$ obtained from the Allan variation the Rb-oscillator. We apply a -0.40(40) kHz correction associated with the measured fractional frequency offset, and assume the frequency instability of the Rb-oscillator and offsets are uncorrelated. We therefore assign a frequency calibration correction of $-0.40(47)$ kHz. 

\subsection{Blackbody radiation}
The blackbody radiation emitted by nearby thermal surfaces perturbs the hydrogen atoms. The off-resonant contribution of the blackbody radiation causes an additional AC Stark shift while the resonant contribution dominantly reduces the lifetime of the states of interest. Numeric calculations of the blackbody shift at 300 K on the 8D states indicate that a shift of $0.49$ kHz is expected  \cite{Farley1981} -- the $2S$ state is unperturbed at our level of precision. The previous measurement of the 2S-8D transitions measured the blackbody shift at 300 K to be $0.65(16)$ kHz by heating the magnetic shields surrounding the spectroscopic volumes by 30 K \cite{DeBeauvoir2000a}, which is in excellent agreement with the theoretical estimation of $0.49$ kHz. We therefore apply a blackbody correction of $-0.49(16)$ kHz.

\subsection{Pressure Systematics}
 The most likely gas species to participate in collisions with the spectroscopy atoms are water and atomic hydrogen. Collisions between a hydrogen atom and a water molecule results in the hydrogen atom experiencing a varying electric field due to the intrinsic dipole moment of the water molecule, which is about 0.72$\,e\,a_0$ \cite{Clough1973}. In contrast, collisions between hydrogen atoms are dominated by the Van der Waals interactions. To estimate potential shifts associated with pressure, we use the impact approximation \cite{Cooper1967, Matveev2019}.

The linewidths of the 2S-8D and 2S-12D transitions obtained in preliminary measurements and at poor vacuum were broader than predicted by up to tens of kHz, with the 12D more strongly effected than the 8D line. We believe this broadening was due to collisions with water molecules. In a collision with a water molecule, hydrogen atoms in the 8D state may be quenched to nearby 8P and 8F states from the varying electric field. 

To quantify possible shifts associated with such collisions, we have employed Monte Carlo simulations of the collisions between hydrogen and water molecules. From these simulations, we find a 16 GHz/Torr broadening and a 3 MHz/Torr shift on the 8D line and 37 GHz/Torr broadening, 3 MHz/Torr shift on the 12D line. 
For pressures in the range of a 10$^{-6}$ Torr, this corresponds to $\sim$ 10 kHz of broadening, which roughly agrees with our observation of broadened lines when we attempted to measure lines with relatively poor vacuum. The associated pressure shift is many order of magnitude below the associated broadening -- 1 kHz of broadening would correspond to a $\sim0.1$ Hz shift. While we do not see evidence of this broadening on our data used in the AC extrapolations, it is difficult to exclude the possibility of 1 kHz broadening on a $\approx 600$ kHz line, and therefore set an upper limit of pressure shifts due to water collisions at 0.1 Hz. 

In principle, the Van der Waals interaction between hydrogen atoms may also induce shifts on the 2S$_{1/2}-$8D$_{5/2}$ line. The Van der Waals interaction energy is given as
\begin{equation}
    H_{V}= \frac{e^2}{4\pi\epsilon_0}\frac{x_1x_2+y_1y_2-2z_1z_2}{R^3},
    \label{VanHam}
\end{equation}
which results in an associated energy shift of
\begin{equation}
    \Delta E_{V}(n)=\sum_{m,m\neq n}\frac{|\bra{n}H_{V}\ket{m}|^2}{E_n-E_m}=\frac{C_6}{R^6},
    \label{Vander perturb}
\end{equation}
with $\ket{n}=\ket{n_1}\otimes\ket{n_2}$ and $\ket{m}=\ket{m_1}\otimes\ket{m_2}$ being product states of atom 1 and atom 2.  Here, we assume atom 1 to be the perturbing atom and atom 2 to be the perturbed atom. Under the impact approximation, the $C_6$ coefficient can be used to estimate the frequency shift due to Van der Waals interactions, which is given by
\begin{equation}
    \Delta\omega_{V}\approx2.9\left(\frac{C_6}{\hbar} \right)^{2/5}v^{3/5}\mathcal{N}_p.
    \label{VDW shift}
\end{equation}
Here, $v$ is the relative velocity between the two hydrogen atoms and $\mathcal{N}_p$ is the number density \cite{Sobelman1972, Beyer2017}. 

The Van der Waals interaction energy may be estimated with
\begin{equation}
    \Delta E_V \sim \frac{1}{4\pi\epsilon_0 R^6}\sumint_{m_1,m_2}\frac{|\bra{n_1,n_2}r_1r_2\ket{m_1,m_2}|^2}{E_{n_1}+E_{n_2}-E_{m_1}-E_{m_2}},
\end{equation}
where the integral-sum symbol indicates that the continuum contribution is also considered.  For clarity, we have not included the angular factors seen in Eqn. \ref{VanHam}, which is justified since we will find the possible systematic shifts associated with these collisions are very small.  We make the substitution $\Delta (m_1)=E_{n_1}-E_{m_1}$, and rearrange the expression slightly to give
\begin{equation}
    \Delta E_V\sim\frac{1}{4\pi\epsilon_0 R^6}\sumint_{m_1}|\bra{n_1}r_1\ket{m_1}|^2\sumint_{m_2}\frac{|\bra{n_2}r_2\ket{m_2}|^2}{E_{n_1}+E_{n_2}+\Delta(m_1)}.
\end{equation}
The sum (including the continuum contribution) over $m_2$ has a complicated closed form expression as a function of the energy difference $\Delta (m_1)$, which we will notate as $S(\Delta(m_1))$ -- for this, we used the Sturmian formalism which may also be used to calculate the matrix elements and AC Stark coefficients for two-photon transitions  \cite{Swainson1991a, Haas2006}. Therefore, calculating the $C_6$ coefficients requires summing the squared dipole moments multiplied by $S(\Delta(m_1))$ over $m_1$. While the sum over the bound states is straightforward, the continuum contribution is more difficult.  As an upper bound for the continuum contribution, we multiply the maximal value of $S(\Delta(m_1))$ in the continuum by the total continuum contribution of the perturbing atom; the continuum contribution for the perturbing $\ket{1S}$ and $\ket{2S}$ hydrogen atoms can be found in \cite{Bethe1957} section 63, table 13 as $0.849 \, a_0^2$ and $2.7 \, a_0^2$ respectively. Additional care must be taken when the perturbing atom is in the $\ket{2S}$ state due to the near degeneracy of the 2S-8P and 2P-8D transitions. For these transitions, we assume the smallest energy splittings between the fine-structure components ($\approx1$ GHz), and the largest dipole matrix element to provide the upper bound. From this, we find
\begin{equation}
    C^{1S}_6\approx5 \, \times \, 10^4 \, hc \, R_\infty \, a_0^6,
\end{equation} 
and 
\begin{equation}
C^{2S}_6\approx3 \, \times \, 10^{11}\, hc\, R_\infty \, a_0^6
\end{equation} 
 with the $C^{2S}_6$ coefficient dominated by the near-degenerate contributions. We estimate an atomic hydrogen density of $10^{13}$/m$^3$ in our spectroscopic volume, with about $1\%$ of those atoms in the metastable state. From Eqn.~(\ref{VDW shift}), this corresponds to a shift of $\sim 1$ Hz from Van der Waals interactions, which is far below our current level of precision. 

\subsection{Cross-damping}

Of particular interest in many of the recent hydrogen measurements is an effect known as cross-damping, or quantum interference \cite{Horbatsch2010,Yost2014,Fleurbaey2017, Beyer2017, Udem2019, Grinin2020}. This effect appears when there are nearby off-resonant transitions and the fluorescence from spontaneous emission of the excited state is detected. This effect is almost entirely absent if all possible decay channels are equally accounted for, or if the remaining metastable population is detected instead of fluorescence \cite{Yost2014,Udem2019}.  Since we are detecting remaining metastable population, our measurement is insensitive to this systematic.

\subsection{8D$_{5/2}$ hyperfine structure}
The hyperfine structure of the 8D$_{5/2}$ state is unresolved and is therefore included into the lineshape fitting function. The splitting of the $F=3$ and $F=2$ manifolds of the 8D$_{5/2}$ state is calculated to be 142.43(14) kHz \cite{Kramida2010}. Modifying the hyperfine splitting terms in the lineshape fitting function by $\pm140$ Hz results in shifts of the fit resonance frequency of $\pm30$ Hz, which we take as the uncertainty.   

\subsection{Photodiode Imperfections}
The photodiode and data acquisition electronics recording the 778 nm cavity transmission has a small voltage offset of about $< 5$ mV that can vary slightly from day to day. This offset shifts the zero-intensity resonance frequency determined by the AC extrapolations. 
The AC Stark shift is about 300 kHz for a transmission PD voltage of 6 V. Therefore, the uncertainty introduced by a variation of 5 mV is about 250 Hz, which is taken as an additional uncertainty. Aside from DC offsets, a nonlinear response in the experimental detector would also introduce a shift to the AC extrapolated frequencies. Therefore, we have also tested the linearity of the experimental detector by comparing with a test detector. The 778 nm power is sent to the experimental detector, with a fraction of the light power picked off for the test detector. This data is then fit with the function
\begin{equation}
    V_{exp}=a_V V_{test}+ b_V V_{test}^2,
\end{equation}
where $a_V$ and $b_V$ are fitting parameters. The ratio $b_V/a_V$ sets the relative strength of the nonlinear response of the photodiode, and we find an upper bound of $b_V/a_V\approx5\times 10^{-4}V^{-1}$. This nonlinearity can be accounted for by adjusting the AC extrapolation fitting function to $\nu_c(V)=\nu_0+a V+b V^3+ a \frac{b_V}{a_V} V^2+\mathcal{O}\left( (b_V/a_V)^2\right)$, and results in shifts up to 0.2 kHz, which we take as an additional uncertainty. We assume the offset and nonlinearity of the photodiode are uncorrelated and assign a net uncertainty of 0.32 kHz for photodiode imperfections.  

\subsection{Incoherent line-pulling}
Nearby atomic resonances can pull the resonance frequency of interest. Transitions to the 8D$_{3/2}$ state are the closest (57.13 MHz) that can contribute. The 2S$_{1/2}-$8D$_{3/2}$ transition has a smaller matrix element than the 2S$_{1/2}-$8D$_{5/2}$ transition, but for this upper bound we assume that they are equal. We estimate this effect by fitting a single Lorentzian to a pair of Lorentzians separated by 57.13 MHz and find a shift of less than 1 Hz, which is negligible.    

\subsection{Light-force shift}
Metastable atoms traversing the 778 nm intensity profile are deflected from straight-line trajectories due to the AC Stark shift of the $\ket{2S}$ state. The energy shift of the $\ket{2S}$ state is given by
\begin{equation}
    \Delta E(2S)=h \, \beta_{ac}(2S) \, I(\vec{r}),
\end{equation}
where $\beta_{ac}$ is the AC Stark shift coefficient \cite{Haas2006}, and $I(\vec{r})$ is the spatially varying intensity seen by the atom.  The nonzero gradient of $I(\vec{r})$ produces a force which can deflect the atom to a region of differing AC Stark shift. We estimate an upper bound for the possible deflection, given as $\delta \vec r$, by considering the atom to be under a constant acceleration, which is found by assuming the maximal possible gradient of $I$ for a given power, during the transit time. This can be estimated as  
\begin{equation}
    \delta\vec{r}\leq\frac{h}{2m}  \, \beta_{ac}(2S) \, \, \nabla I|_{\text{max}} \, \, (\Delta t)^2.
\end{equation} 
For 50 W in a Gaussian beam with a beam diameter of 600 $\mu$m, this corresponds to a maximal deflection of $\sim 10$ nm. This can correspond to an intensity variation of 10 ppm at most, which could vary the resulting AC Stark shift by a few Hz at the largest intensities. This effect is negligible at our current level of precision.   

\section{Effect of an additional Yukawa Potential on the Rydberg Constant Extractions}
Here we discuss the effect of an additional perturbative potential of the form 

\begin{equation}
V(r)=\beta \,  \frac{e^{-r/\alpha}}{r},
\end{equation}
on the extraction of the Rydberg constant in hydrogen spectroscopy \cite{Jones2019}. Here, $\beta$ is the effective strength of the potential, and $\alpha$ gives the length scale. Following the discussion in \cite{Jones2019}, we allow for the sign of $\beta$ to be both positive and negative.  

\begin{figure} [h!]
    \centering
    \includegraphics[width=.6\linewidth]{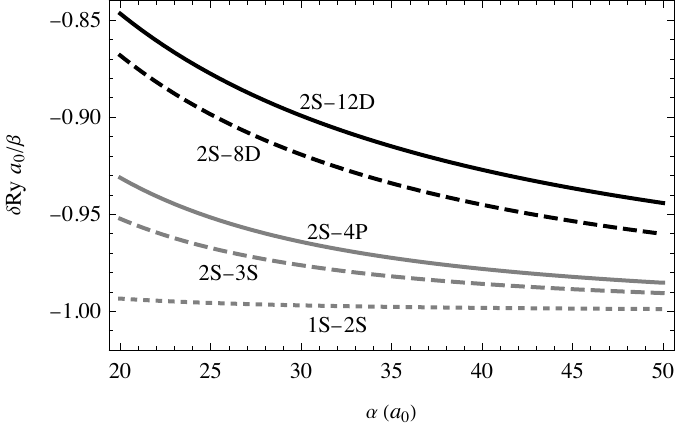}
    \caption{Change in the extracted Rydberg energy due to a perturbing Yukawa potential (in units of $\beta/a_0$) as a function of the length scale $\alpha$.  1S-2S: dotted grey, 2S-3S: dashed grey,  2S-4P: solid grey, 2S-8D: dashed black, 2S-12D solid black.}
    \label{RydbergPlot}
\end{figure}

\begin{figure} [b!]
    \centering
    \includegraphics[width=.6\linewidth]{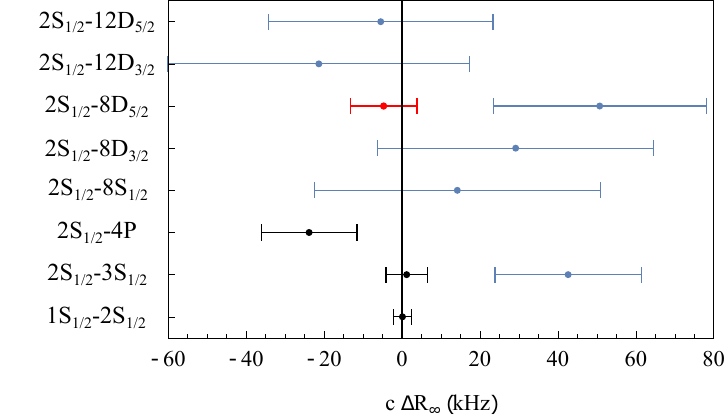}
    \caption{Variation of Rydberg constant extractions from the weighted mean for the measured transitions shown in Fig.~5 of the main text but with the addition of the Yukawa potential perturbation. For this example, the length scale of the Yukawa potential is $\alpha=$34 $a_0$ and $\beta= 1.64 \times 10^{-10}\, Ry \, a_0$. Blue points were measured in Paris (2S-3S \cite{Fleurbaey2018},  2S-8S/D \cite{DeBeauvoir1997}, 2S-12D \cite{DeBeauvoir2000a}), black points were measured in Garching (1S$_{1/2}$-2S$_{1/2}$ \cite{Parthey2011}, 2S$_{1/2}$-3S$_{1/2}$ \cite{Grinin2020}, 2S-4P \cite{Beyer2017}), and our result is in red.}
    \label{RydbergReducedChi2}
\end{figure}
We use first-order perturbation theory to calculate the effect of this perturbing potential on the hydrogen energy levels, given by $\delta E(n,l)$. The effect of this perturbation on an extraction of the Rydberg energy, $Ry=h c R_\infty$, from a specific measured transition is then given by
\begin{equation}
    \frac{\delta Ry}{Ry}=\frac{\delta E(n_2,l_2)-\delta E(n_1,l_1)}{E(n_2,l_2)-E(n_1,l_1)},
\end{equation}
where $E(n,l)$ is the energy of an unperturbed energy level and the subscripts 2 and 1 indicate the upper and lower states respectively, and $\delta Ry$ is the change in the Rydberg energy determination. In Fig.~\ref{RydbergPlot}, we show, $\delta Ry$, scaled by $\beta/a_0$ as a function of the length scale $\alpha$. As can be seen, the perturbation generally gives a larger Rydberg constant extraction as $n_2$ and $n_1$ increase provided that $\beta$ is positive.

By considering an example  length scale of $\alpha=34 \, a_0$, where $a_0$ is the Bohr radius, and by scaling the effective strength of the potential to $\beta = 1.64 \times 10^{-10}\, Ry \, a_0$, the tension in the extractions of the Rydberg constant through hydrogen laser spectroscopy can be significantly reduced.  This result is shown in Fig.~\ref{RydbergReducedChi2}. The extractions with this example Yukawa potential perturbation have a reduced chi-squared statistic of 2.0 while the data shown in Fig. 5 of the main text has a reduced chi-squared statistic of 4.0.